\documentclass[a4paper]{jpconf}
\usepackage{graphicx}
\begin{document}
\title{Fragmentation functions for pions, kaons, protons and charged hadrons}

\author{D de Florian$^1$, R Sassot$^1$ and M Stratmann$^2$}

\address{$^1$ Departamento de Fisica, Universidad de Buenos Aires, 
Ciudad Universitaria, Pab.\ 1 (1428) Buenos Aires, Argentina}
\address{$^2$ Radiation Laboratory, RIKEN, 2-1 Hirosawa, Wako, Saitama 
351-0198, Japan}

\ead{deflo@df.uba.ar, sassot@df.uba.ar, marco@ribf.riken.jp}

\begin{abstract}
We present new sets of pion, kaon, proton and inclusive charged hadron 
fragmentation functions obtained in NLO combined analyses of single-inclusive 
hadron production in electron-positron annihilation, proton-proton collisions, 
and deep-inelastic lepton-proton scattering with either particles identified 
in the final state. At variance with all previous fits, the present analyses 
take into account data where hadrons of different electrical charge are 
identified, which allow to discriminate quark from anti-quark fragmentation 
functions. 
\end{abstract}

\section{Introduction}
In the last few years there has a growing interest in accurate 
parameterizations for fragmentation functions driven by the increasing role
of 
one particle inclusive measurements as tools able to provide 
valuable information on many subjects, including the spin and flavor 
structure of the nucleon, nuclear modifications of parton densities 
and fragmentation functions, and in general, as a window 
to the non-perturbative regime, much more incisive than totally 
inclusive measurements.

In this context, parametrizations for fragmentation functions have been 
extracted, however, mostly based on single inclusive electron-positron 
annihilation measurements. These data give per se no information on how to 
disentangle quark from anti-quark fragmentation, and fixes mainly the flavor 
singlet combinations of fragmentation functions at intermediate hadron energy 
fractions. The gluon fragmentation is 
also not exceedingly well constrained, since its contribution enter as a 
higher order correction, and the scale dependence is too weak to determine 
it.

In the following we very briefly summarize results from a global analysis where
we have determined individual fragmentation functions for quark and 
anti-quarks for all flavors, as well as gluons, from a much larger set of 
data \cite{pika,hppbar}. The addition of semi-inclusive deep inelastic 
scattering (SIDIS) and proton-proton collisions 
measurements not only increases statistics, but have the advantages of charge 
discriminated final states, different ranges for the 
scale and energy fractions, and different sensitivity to the partonic species.

\section{Parameterizations and results}
In order to have the flexibility required by charge separated distributions
and to accommodate the additional data, we adopt a somewhat more versatile 
functional form for the input distributions than in previous extractions of
fragmentation functions
\cite{bourhis,kretzer,akk,hirai}
\begin{equation}
\label{eq:ff-input}
D_i^H(z,\mu_0) =
{N_i z^{\alpha_i}(1-z)^{\beta_i} [1+\gamma_i (1-z)^{\delta_i}] },
\end{equation}
where the initial scale $\mu_0$ in Eq.~(\ref{eq:ff-input}) is taken to be 
$\mu_0=1\,\mathrm{GeV}$ for the lighter partons and the quark masses for the 
heavier ones. To reduce the number of parameters to those that can be 
effectively constrained by the data, we are forced to make some plausible
assumptions; for example we impose isospin symmetry for the 
sea fragmentation functions in the case of pions, i.e. 
$D_{\bar{u}}^{\pi^+}=D_{d}^{\pi^+}$, but we allow for slightly 
different normalizations in the $q+\bar{q}$ sum:
$D_{d+\bar{d}}^{\pi^+}= N D_{u+\bar{u}}^{\pi^+}$.
For strange quarks it is assumed that 
$D_s^{\pi^+}=D_{\bar{s}}^{\pi^+} =N^{\prime} D_{\bar{u}}^{\pi^+}$. Similar assumptions are made in the case of kaons, protons, and the remaining charged hadrons. For further details see \cite{pika,hppbar}.



\begin{figure}[h]
\begin{minipage}{18pc}
\includegraphics[width=18pc]{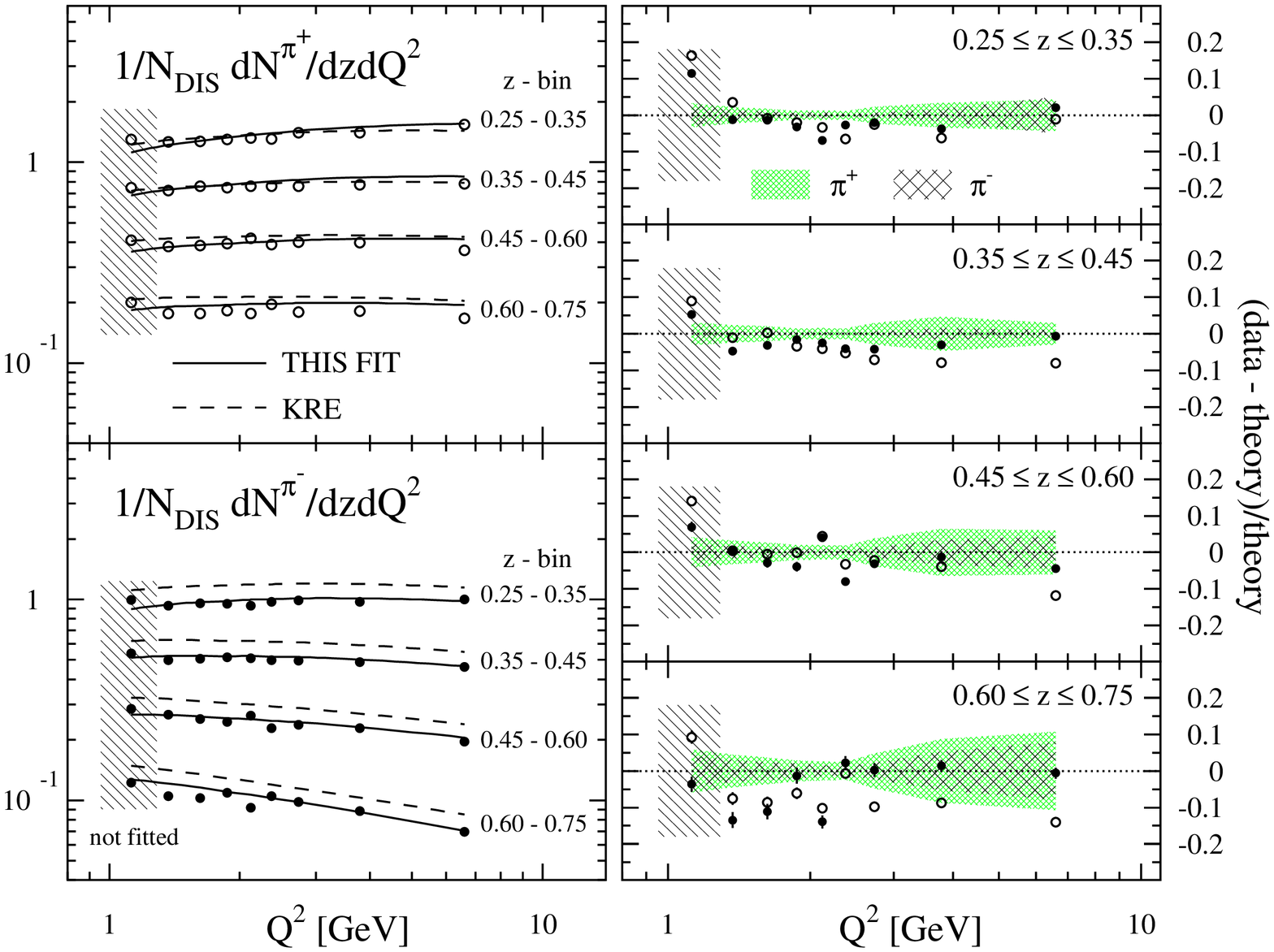}
\caption{\label{label1}Comparison with HERMES pion SIDIS data \cite{data}. 
The label KRE denotes estimates with the set of reference \cite{kretzer}}
\end{minipage}\hspace{1pc}%
\begin{minipage}{18pc}
\includegraphics[width=18pc]{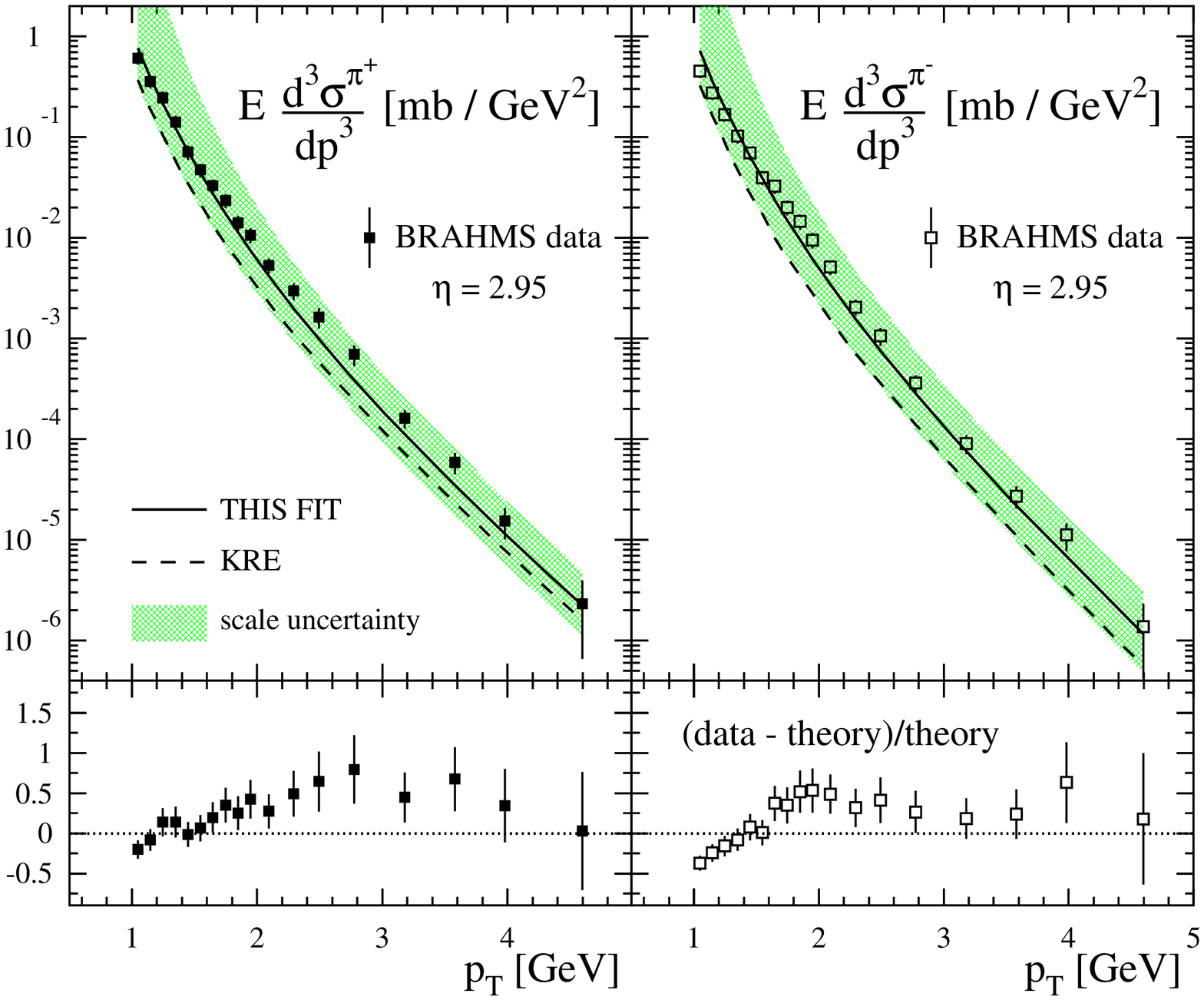}
\caption{\label{label2}Comparison with BRAHMS pion production data in pp 
collisions \cite{data}.}
\end{minipage} 
\end{figure}
\begin{figure}[h]
\begin{minipage}{18pc}
\includegraphics[width=18pc]{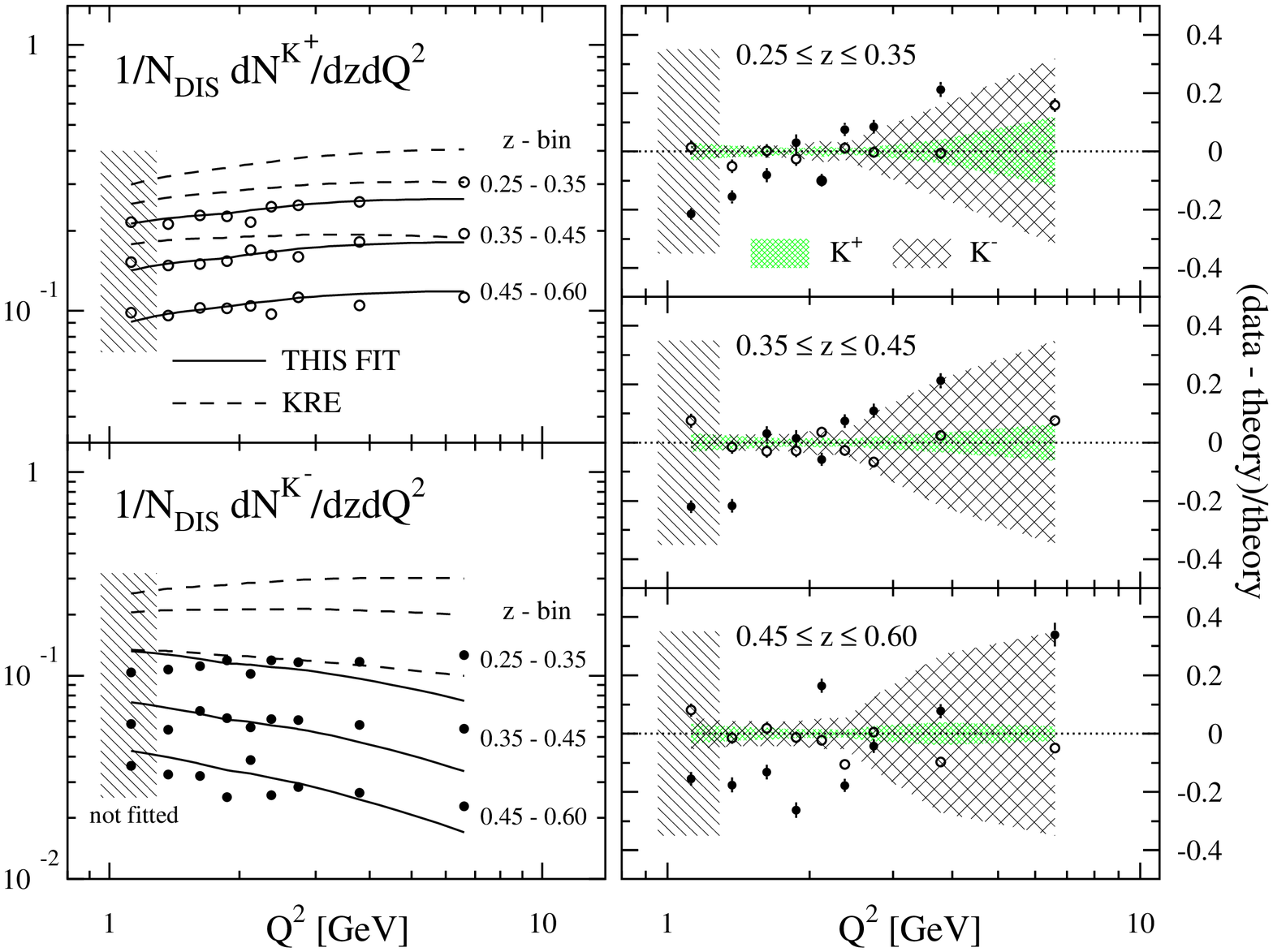}
\caption{\label{label3}The same as Fig.1 but for kaons.}
\end{minipage}\hspace{1pc}%
\begin{minipage}{18pc}
\includegraphics[width=18pc]{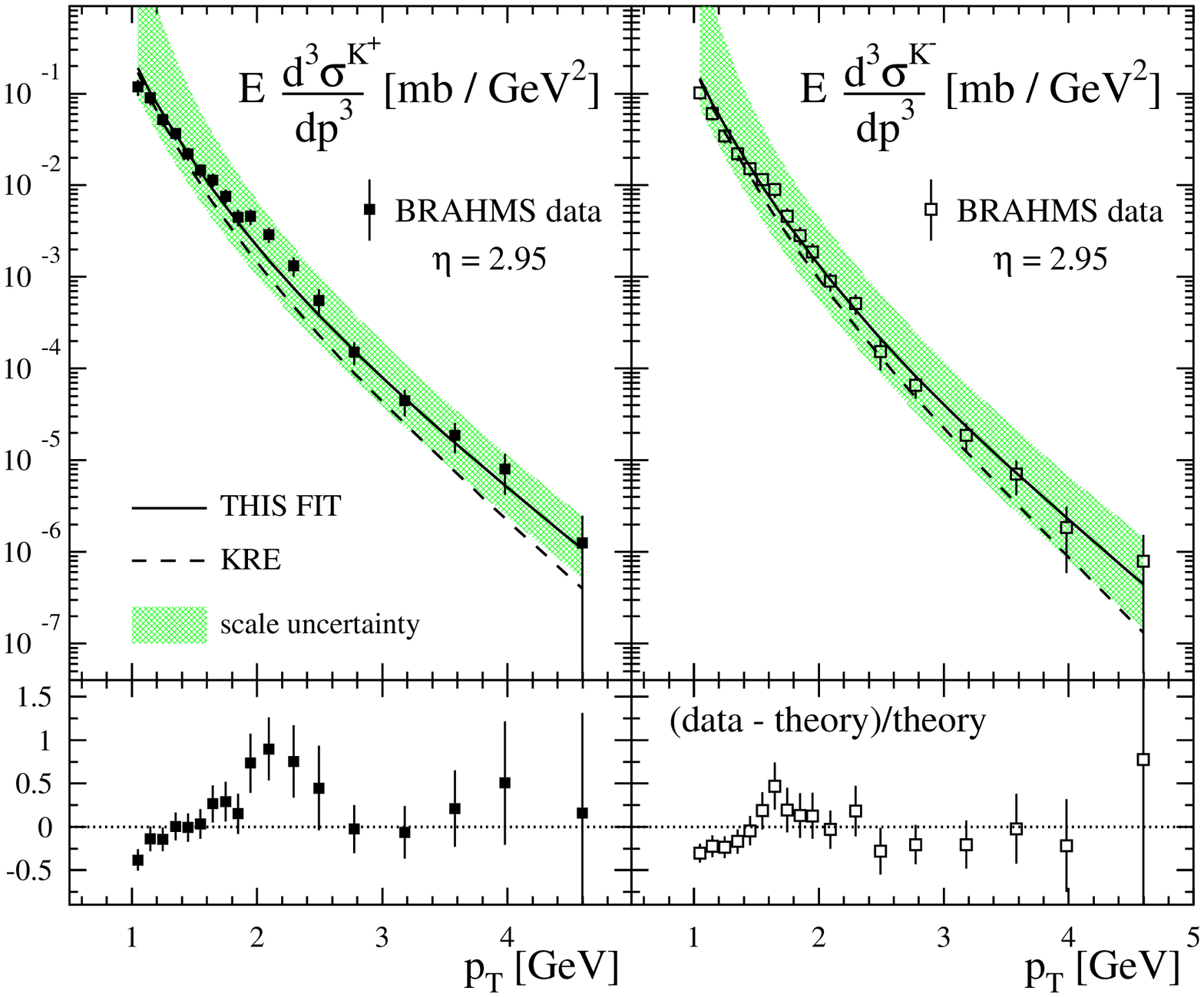}
\caption{\label{label4}The same as Fig.2 but for kaons.}
\end{minipage} 
\end{figure}
As a common feature of all the fits, we find that our new sets of 
fragmentation functions are hardly to distinguish from the previous 
ones when compared to electron positron annihilation data, but as expected,
the differences become apparent when they are compared to data that imply
charge separation (Figs. 1-4), specially in the case of kaons (Figs. 3, 4),
and for observables sensitive to the gluon fragmentation and large hadron 
energy fractions (Fig. 5). 
\begin{figure}[h]
\begin{minipage}{12pc}
\includegraphics[width=12pc]{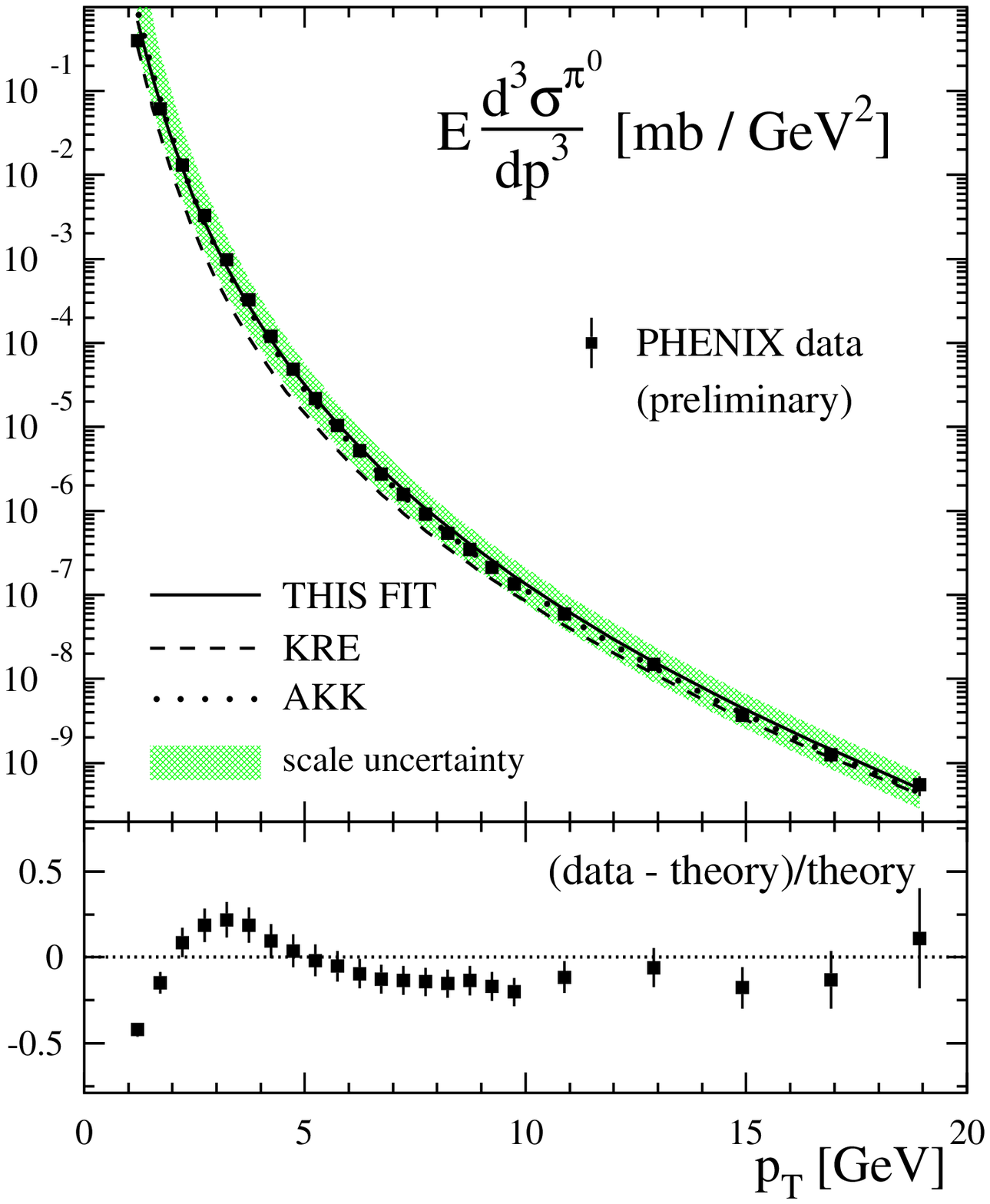}
\end{minipage}\hspace{1pc}%
\begin{minipage}{12pc}
\includegraphics[width=12pc]{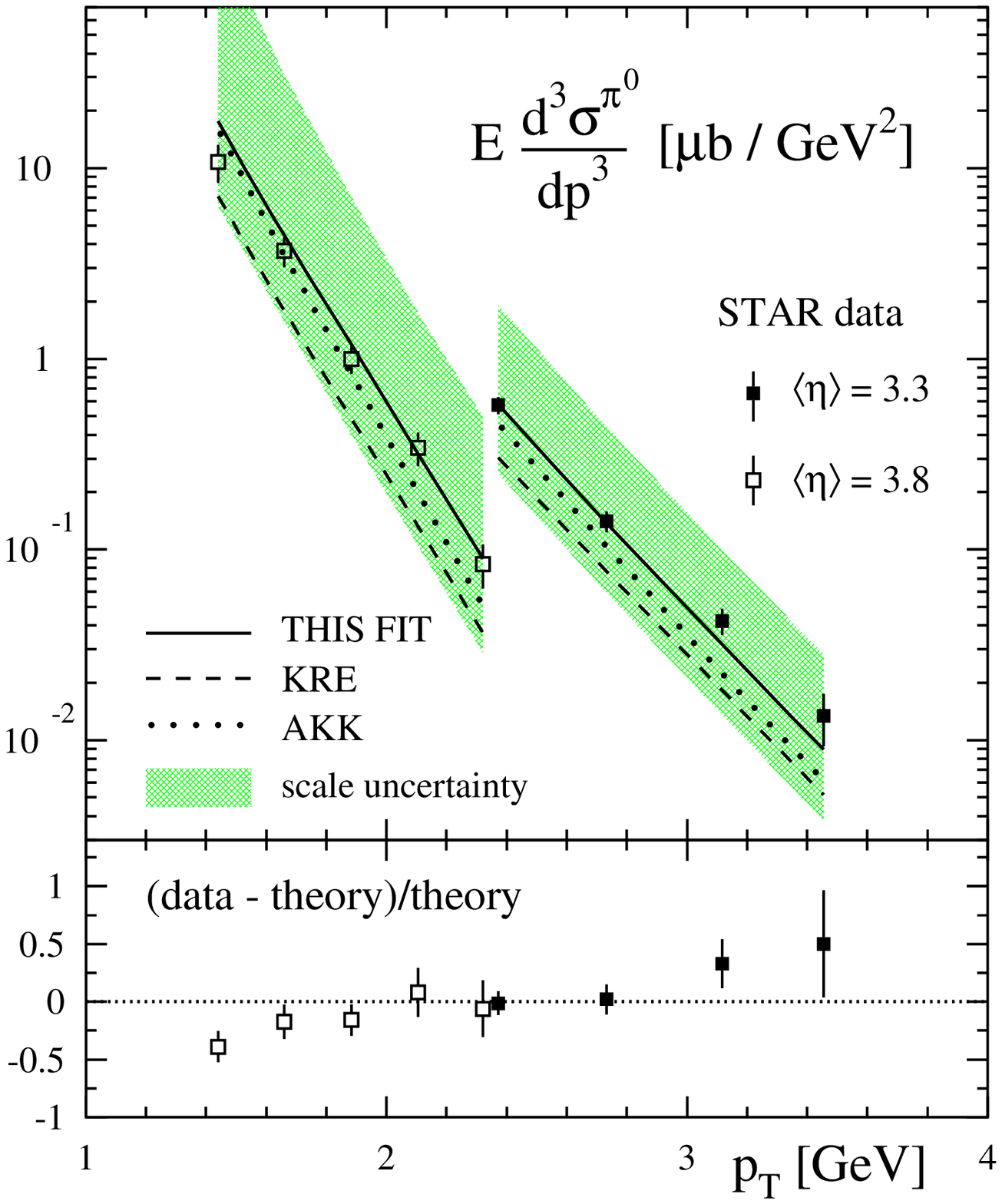}
\end{minipage} 
\begin{minipage}{12pc}
\includegraphics[width=12pc]{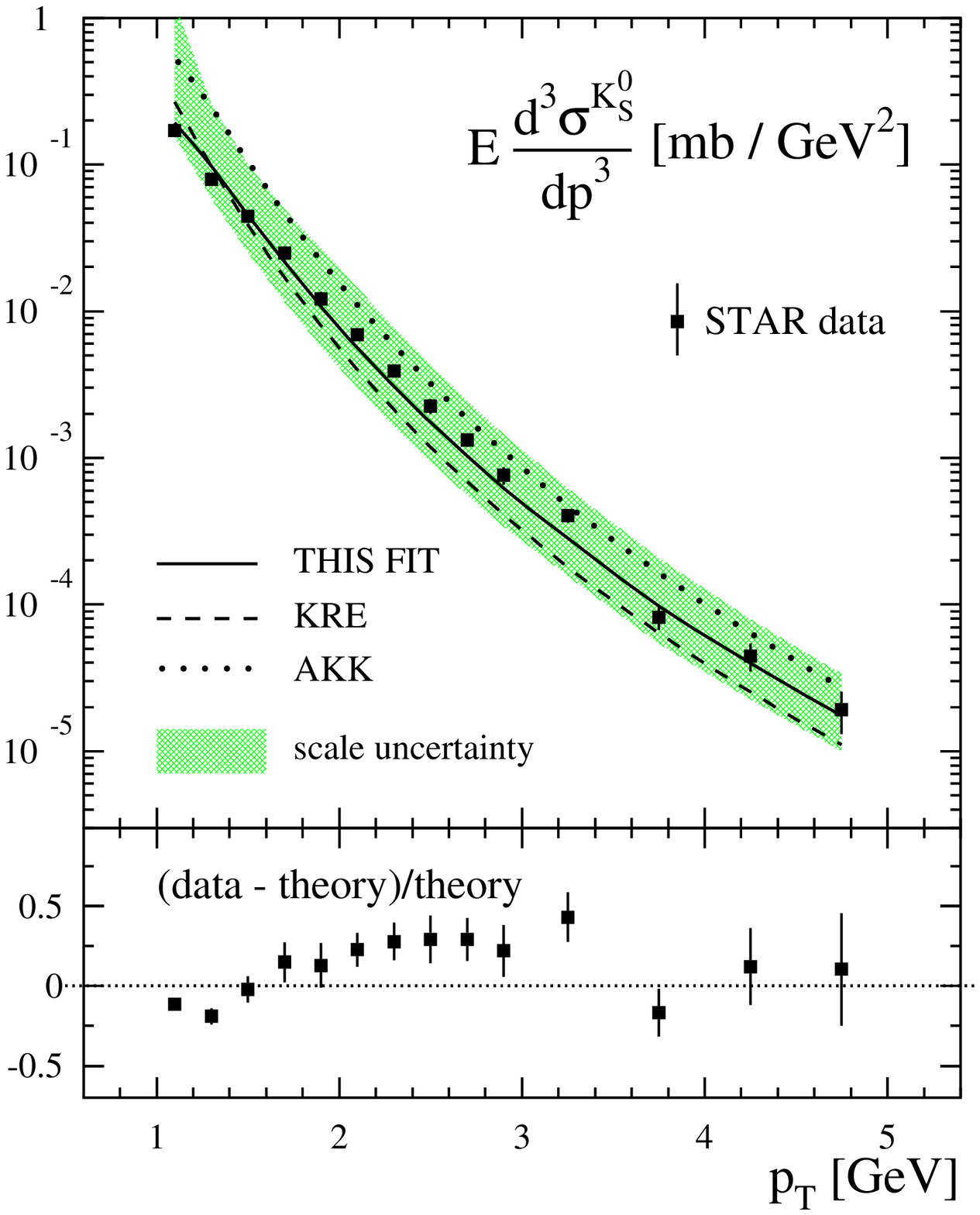}
\end{minipage} 
\caption{\label{label5}Comparison with PHENIX and STAR neutral pion and kaon 
production data in pp collisions \cite{data}. The label KRE and AKK denotes 
estimates with the sets in \cite{kretzer} and  \cite{akk} 
respectively.}
\end{figure}
Proton as well as inclusive charged hadron production data are well reproduced
(Figs. 6, 7). Fortran routines providing LO and NLO fragmentation functions 
are available upon request from the authors. 
\begin{figure}[h]
\begin{minipage}{18pc}
\includegraphics[width=18pc]{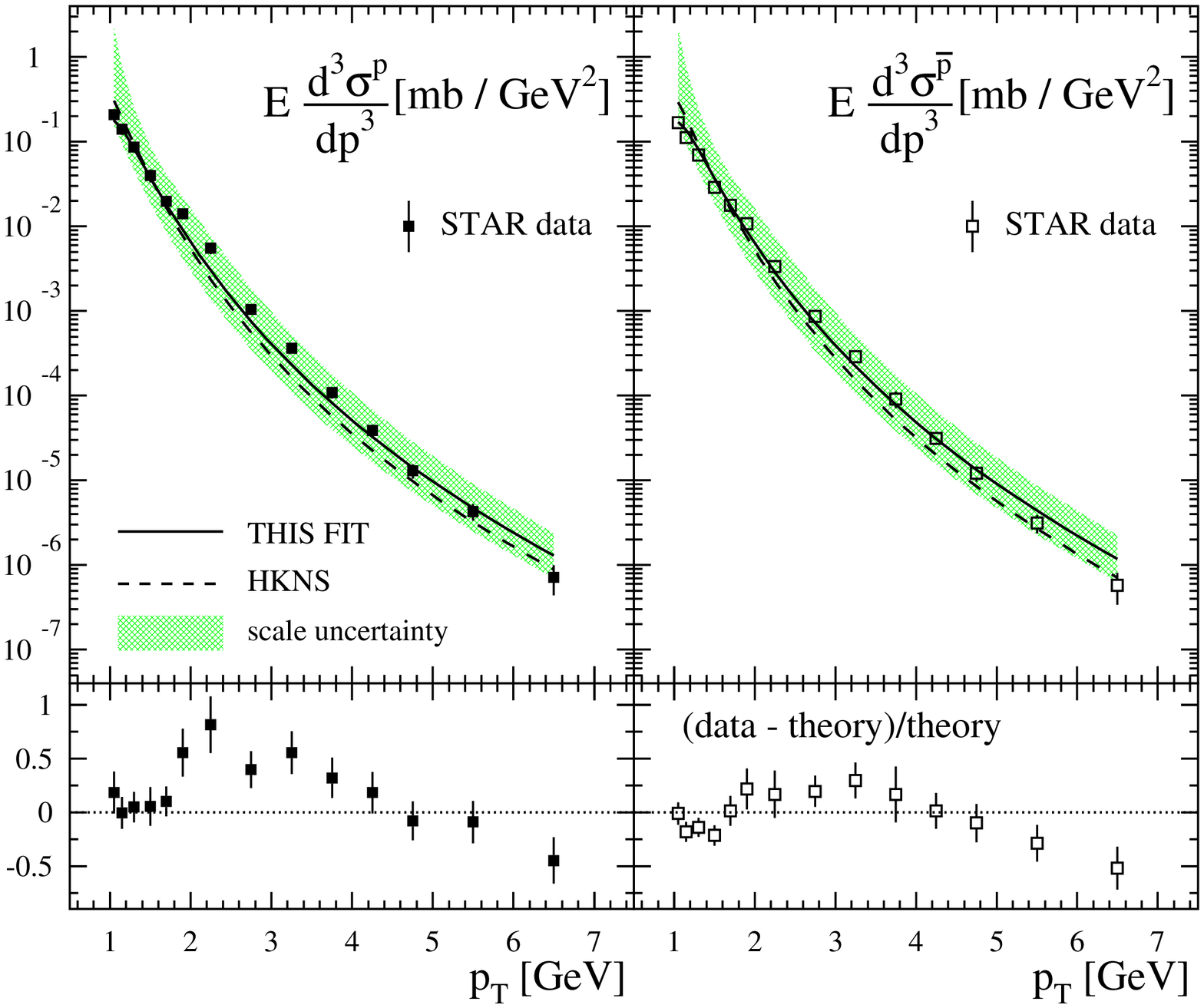}
\caption{\label{label6}Comparison with STAR data on (anti)proton production 
\cite{data} and the set in \cite{hirai}.}
\end{minipage}\hspace{1pc}%
\begin{minipage}{18pc}
\includegraphics[width=12.5pc]{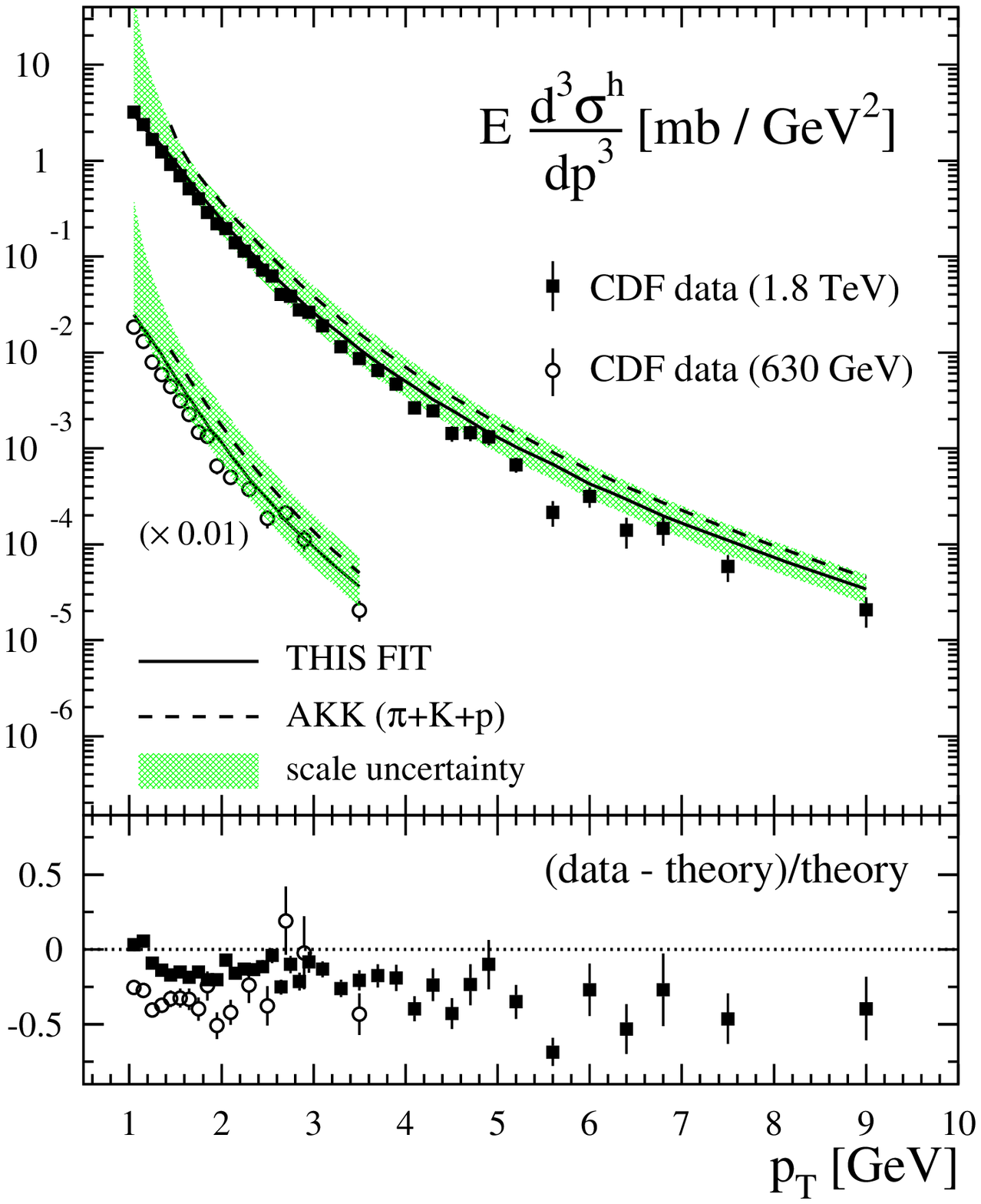}
\caption{\label{label7}Comparison with CDF inclusive charged hadron production data \cite{data}}
\end{minipage} 
\end{figure}



\section*{References}
\begin{thebibliography}{9}
\bibitem{pika}de Florian D, Sassot R and Stratmann M 2007 {\it Phys. Rev.} D 
{\bf 75} 114010 
 \bibitem{hppbar}de Florian D, Sassot R and Stratmann M 2007  Global analysis of fragmentation functions for protons and charged hadrons {\it Preprint} 
arXiv:0707:1506 [hep-ph]
\bibitem{bourhis}
Bourhis L, Fontannaz M, Guillet J, and Werlen M 2001
{\it Eur. Phys. J.} C {\bf 19} 89
\bibitem{kretzer} Kretzer S 2000 {\it Phys. Rev.} D {\bf 62} 054001 
\bibitem{akk} Albino S, Kniehl B A and Kramer G 2005
{\it Nucl. Phys.} B {\bf 725} 181
\bibitem{hirai} Hirai M, Kumano S, Nagai T H and Sudoh K 2007 {\it Phys. Rev.}
D {\bf 75} 094009 
\bibitem{data} For a detailed account of the data sets included in the
fit see [1] and [2]. 


\end{thebibliography}
\end{document}